\documentclass[10pt,twocolumn,letterpaper]{article}

\usepackage{cvpr}
\usepackage{times}
\usepackage{epsfig}
\usepackage{graphicx}
\usepackage{amsmath}
\usepackage{amssymb}

\usepackage[breaklinks=true,bookmarks=false]{hyperref}

\cvprfinalcopy % *** Uncomment this line for the final submission

\def\cvprPaperID{****} % *** Enter the Workshop Paper ID here

\begin{document}

%%%%%%%%% TITLE
\title{Multi-scale Grouped Dense Network for VVC Intra Coding}

\author{Xin Li, Simeng Sun \thanks{Equal Contribution}, Zhizheng Zhang, Zhibo Chen \thanks{Corresponding Author}\\
	CAS Key Laboratory of Technology in Geo-spatial Information Processing and Application System\\
	University of Science and Technology of China\\
	{\tt\small \{lixin666, smsun20, zhizheng\} @mail.ustc.edu.cn, chenzhibo@ustc.edu.cn}
}

\maketitle
\pagestyle{empty}
\thispagestyle{empty} 
%\thispagestyle{empty}

%%%%%%%%% ABSTRACT
\begin{abstract}
   Versatile Video Coding (H.266/VVC) standard achieves better image quality when keeping the same bits than any other conventional image codec, such as BPG, JPEG, and etc. 
   However, it is still attractive and challenging to improve the image quality with high compression ratio on the basis of traditional coding techniques. In this paper, we design the multi-scale grouped dense network (MSGDN) to further reduce the compression artifacts by combining the multi-scale and grouped dense block, which are integrated as the post-process network of VVC intra coding. Besides, to improve the subjective quality of compressed image, we also present a generative adversarial network (MSGDN-GAN) by utilizing our MSGDN as generator. Across the extensive experiments on validation set, our MSGDN trained by MSE losses yields the PSNR of 32.622 on average with teams "IMC" and "haha" at the bit-rate of 0.15 in Low-rate track. Moreover, our MSGDN-GAN could achieve the better subjective performance.
\end{abstract}

%%%%%%%%% BODY TEXT
\section{Introduction}
Image/video compression algorithms such as JPGE, JPEG2000 and BPG are of practical importance for multimedia data storage and transmission.
%have achieved great development. Traditional image compression algorithms such as JPGE \cite{wallace1992jpeg} and JPEG2000 \cite{christopoulos2000jpeg2000}, and BPG have been widely used for image storage, transmission and so on. 
As the next generation of video coding technology, Versatile Video Coding (H.266/VVC) standard are developed by Video Coding Experts Group (VCEG) of ITU-I and Moving Picture Experts Group (MPEG) of ISO/IEC, which achieves up to 30$\%$ bit-rate reduction compared with H.265/HEVC. However, various coding artifacts, such as blocking and blurring, are still serious when the compression ratio gets higher. To further eliminate compression artifacts, some CNN-based filtering methods \cite{wang2019integrated, wang2019attention} have been used in in-loop filters of VVC. 
\begin{figure}[htp]
	\centering
	\includegraphics[width=\linewidth]{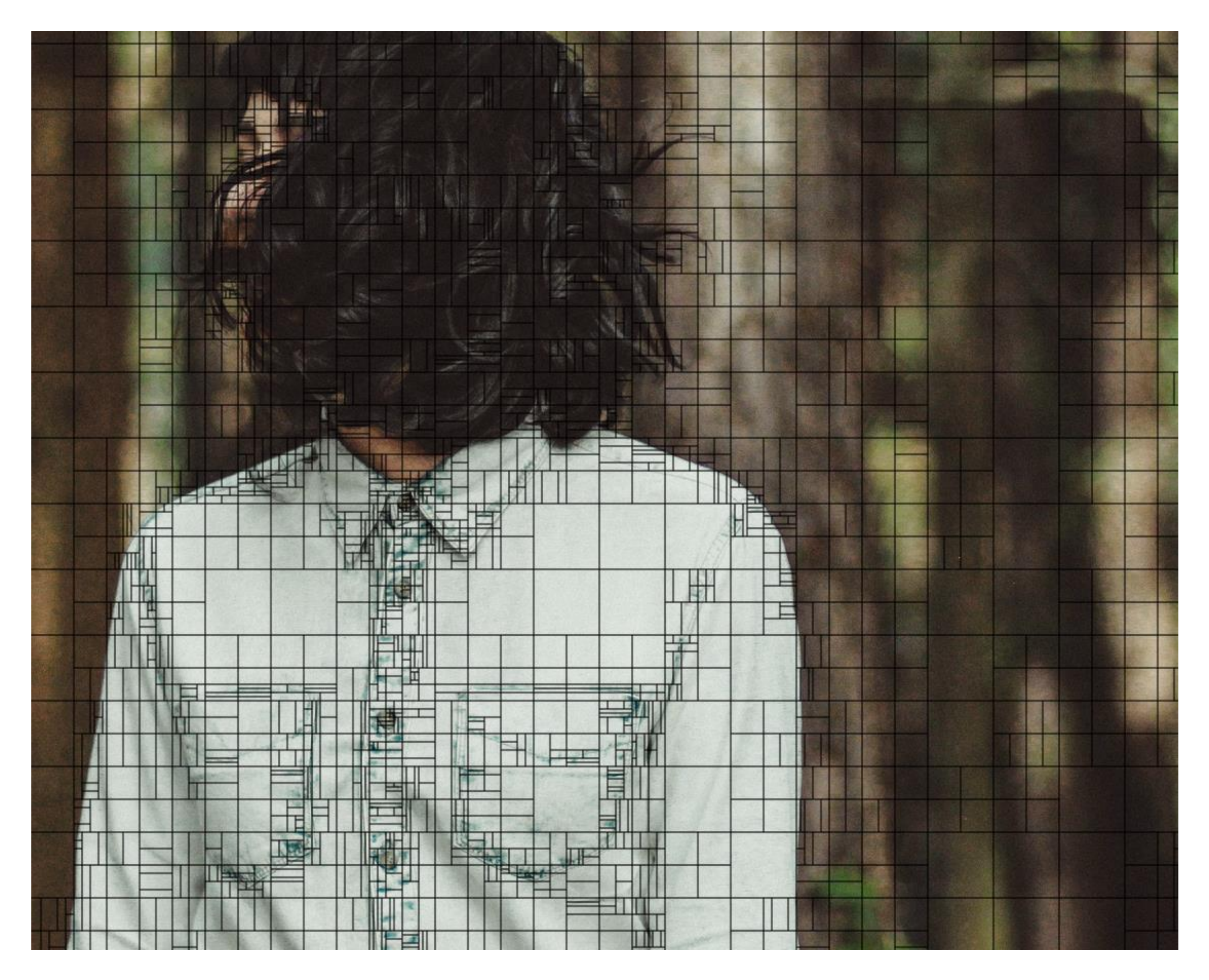}
	\caption{Example of CTU partition in VVC.}
	\label{fig:partition}
\end{figure}

With the development of deep neural network, learning based image compression algorithms \cite{balle2016end, lee2018context, lee2019hybrid, chen2019learning,zhang2019learned,he2019beyond, jin2018multiscale} have drawn huge attention. Belie $et$ $al.$ \cite{balle2016end} implemented the end-to-end optimized image compression by utilizing the general divisive normalization and soft quantization with additive uniform noise. To further optimize the entropy coding, Lee $et$ $al.$ \cite{lee2018context} proposed the context adaptive entropy model to further improve the compression efficiency, which exceeded BPG. Moreover, with the image compression and image enhancement network in cascade, Lee $et$ $al.$ \cite{lee2019hybrid} achieved the comparable performance to VVC intra coding. Despite it, learning based codecs are not very efficient as conventional codecs in their computational complexity. Hence, in this paper, we propose a learning based approach as a post-process technique to improve the performance of VVC intra coding.
%Nevertheless, it still requires some time to promote the learning based image coding technologies due to the limited devices. In this paper, we focus on the improvement of VVC intra coding.    
\begin{figure*}[htp]
	\centering
	\includegraphics[width=0.9\linewidth]{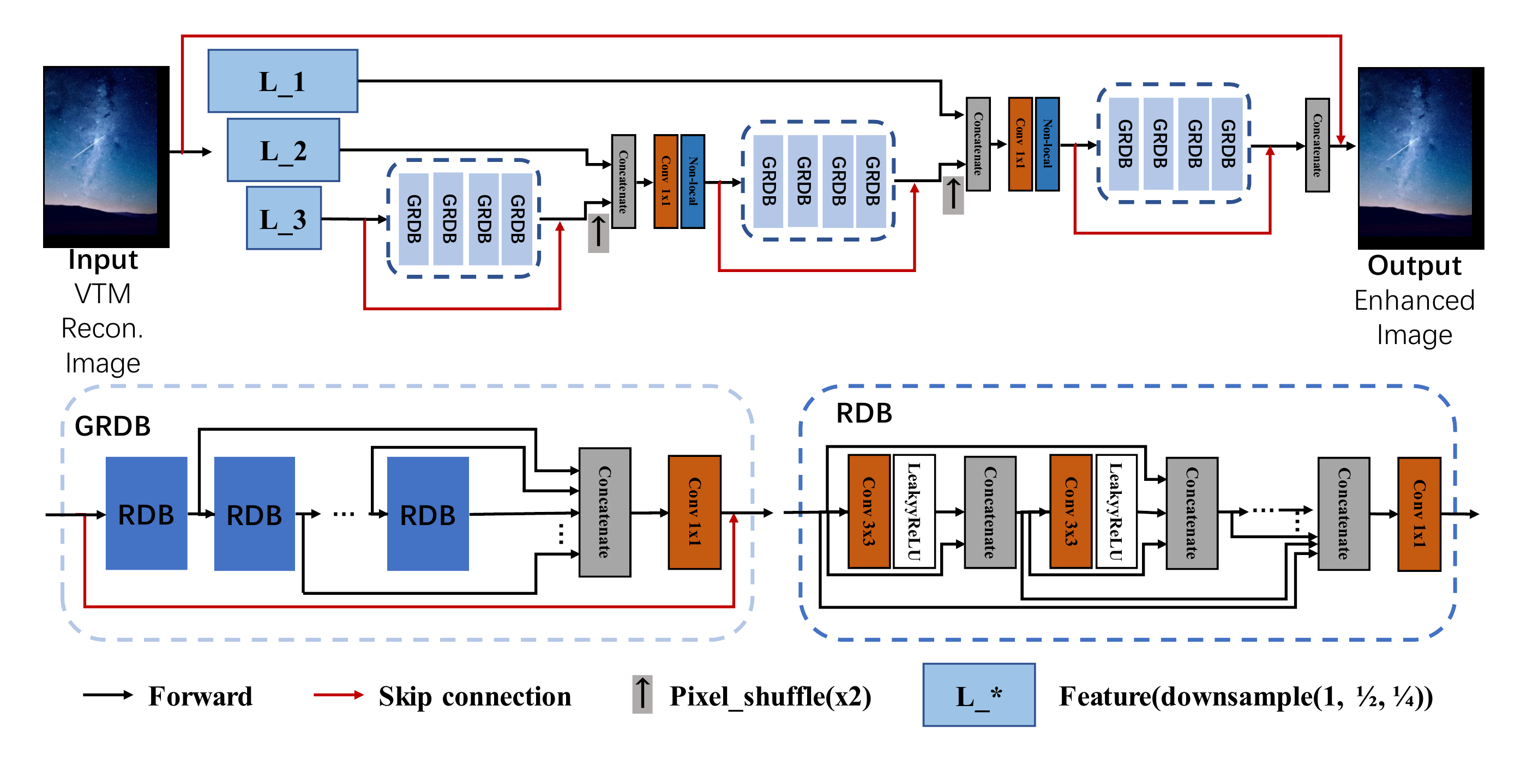}
	\caption{Overview of our Multi-scale grouped dense network (MSGDN).}
	\label{fig:framework}
\end{figure*}

To this regard, some studies \cite{xue2019attention, lu2019learned} have applied the network of image restoration to the post process of compression frameworks. However, these methods are not designed according to the distortion characteristics of VVC.
As shown in Fig. \ref{fig:partition}, the size of coding units in VVC are different, which caused the compression artifacts distributed in different spatial size. Moreover, the compression artifacts including blurring, blocking and texture loss are complex, which require the network to have strong representation ability. Therefore, in this paper, we propose the multi-scale grouped dense network (MSGDN) as the post-process of VVC intra coding based on these analysis.

Specifically, inspired by \cite{lu2019learned}, we utilize multi-scale architecture to remove the compression artifacts distributed in different spatial sizes. Moreover, to increase the representation ability of network, we utilize the GRDB \cite{kim2019grdn} as backbone in each scale. By combining the multi-scale and grouped residual dense block, we propose a post-process network named MSGDN, which achieves the state-of-the-art compression performance with the same training datasets. Moreover, to further improve the subjective quality of the compressed image at low bit-rate, we also propose a generative adversarial network (MSGDN-GAN) by utilizing our MSGDN as generator. 

We conduct extensive experiments on validation sets and our MSGDN trained by MSE losses achieves the highest PSNR of 32.62 on average at bit-rate of 0.15. Moreover, our MSGDN-GAN could achieve the better subjective performance as shown in Fig. \ref{fig:compare_3}. 

\section{Approaches}

\subsection{Multi-scale grouped dense network}
 Across different scaled feature, the network could fuse the coarse and fine information to enhance the representational ability. To process the compression artifacts across from different spatial dimensions, we utilize the multi-scale feature extraction as Fig. \ref{fig:framework}. We apply the convolution layer with stride 2 to implement the down-sample of feature. The multi-scale features are composed of three layers. The resolution of first layer is the same as original image and other two layers are down-sampled by 2 and 4 respectively. To enhance the representation power of each layer, we utilize the GRDB \cite{kim2019grdn} in each layer. 
%GRDB is proposed by \cite{kim2019grdn}, which have been employed in the NITRE2019 Real Image Denoising Challenge – Track 2. 
The channels of GRDB in three layers from low to high are set as 128, 128, and 64. Each GRDB contains 4 RDBs and each RDB contains 8 convolution layers. To fusion multi-scale feature representation, we first up-sample the low representation and concatenate it with higher representation followed by one 1x1 convolution layer and non-local module.

\subsection{Generative adversarial network for VVC}
Generative adversarial network has been used in image restoration,  image-to-image translation and style transfer, which could improve subjective quality efficiently. Inspired by \cite{wang2018esrgan}, we utilize our MSGDN as generator. Then we enhance the discriminator with Relativistic average Discriminator (RaD) \cite{jolicoeur2018relativistic}, which focus on whether input data are more relative realistic or not.
% which predicts the probability that input data is more realistic than a randomly sampled data of the opposing type rather than estimates the probability that one image is real. 
\begin{figure}[htp]
	\centering
	\includegraphics[width=0.95\linewidth]{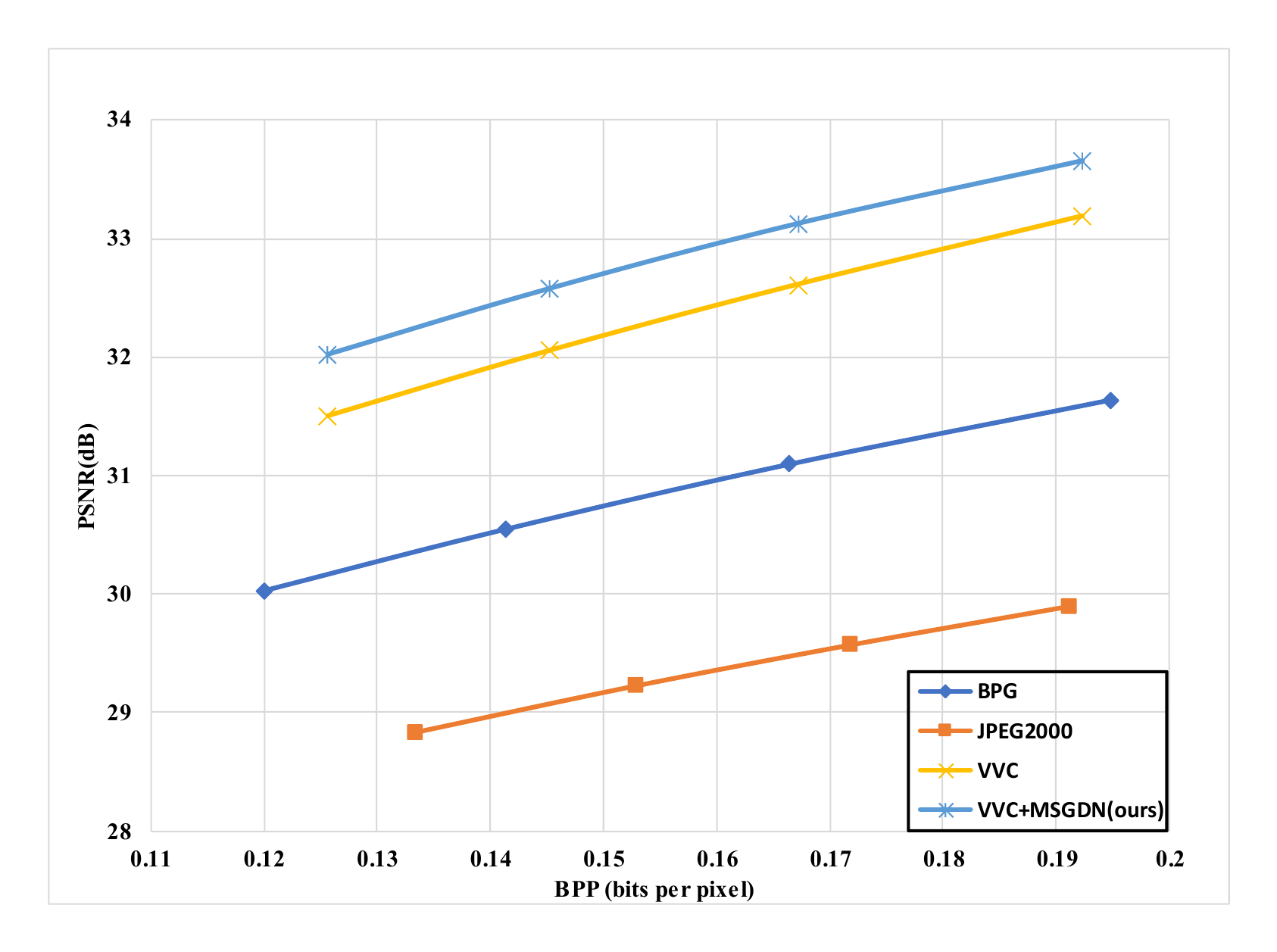}
	\caption{Compression performance on validation sets, compared with JPEG2000, BPG and VVC.}
	\label{fig:compare_1}
\end{figure}

Here, we formula real image and fake image as ${x}_{r}$ and ${x}_{f}$ respectively. The discriminator loss could be formulated as Equ. \ref{lossd}.

\begin{equation}
\label{lossd}
\begin{aligned}
L_{D}^{Ra}&=-{{E}_{{{x}_{r}}}}[\log (D({{x}_{r}},{{x}_{f}}))]\\
&-{{E}_{{{x}_{f}}}}[\log (1-D({{x}_{f}},{{x}_{r}}))],
\end{aligned}
\end{equation}
where ${E}_{{{x}_{r}}}$ means the average of all real data in the mini-batch. $D({{x}_{f}},{{x}_{r}})$ is defined in \cite{jolicoeur2018relativistic} as Equ. \ref{rad}.  
\begin{equation}
D({{x}_{r}},{{x}_{f}})=f(C({{x}_{r}})-{{E}_{{{x}_{f}}}}[C({{x}_{f}})]),
\label{rad}
\end{equation}
where $f$ is sigmoid function and $C$ represents the output of discriminator.
We also adopt perceptual loss in \cite{wang2018esrgan} based on the convolution layer of VGG-19 to further constrain the subjective performance. As shown in Fig. \ref{fig:compare_3}, our MSGDN-GAN achieves the better performance compared with traditional compression methods. However, it will cause the reduction of objective performance.

\subsection{Loss function}
We adopt different loss function for different optimization goals. To improve objective performance, we first utilize the L1 loss as our loss function. Then we fine-tune the MSGDN with MSE loss. In this way, we achieve the highest PSNR of 32.622 on validation set. On the other hand, in order to improve the subjective performance, we first train our generator with L1 loss, and then we replace the L1 loss with a hybrid loss, which is composed of L1 loss, adversarial loss and perceptual loss as Equ. \ref{loss}.
\begin{equation}
\label{loss}
Loss=L1loss+0.01{{L}_{adv}}+0.0001{{L}_{p}},
\end{equation}
where ${L}_{adv}$ represents the adversarial loss and ${L}_{p}$ represents the perceptual loss. ${L}_{adv}$ could be computed with Equ. \ref{lossadv}.
\begin{equation}
\label{lossadv}
\begin{aligned}
L_{adv}&=-{{E}_{{{x}_{r}}}}[\log (1-D({{x}_{r}},{{x}_{f}}))]\\
&-{{E}_{{{x}_{f}}}}[\log (D({{x}_{f}},{{x}_{r}}))],
\end{aligned}
\end{equation}
which is symmetrical form of Equ. \ref{lossd}.

\section{Experiment}

\begin{figure}[htp]
	\centering
	\includegraphics[width=\linewidth]{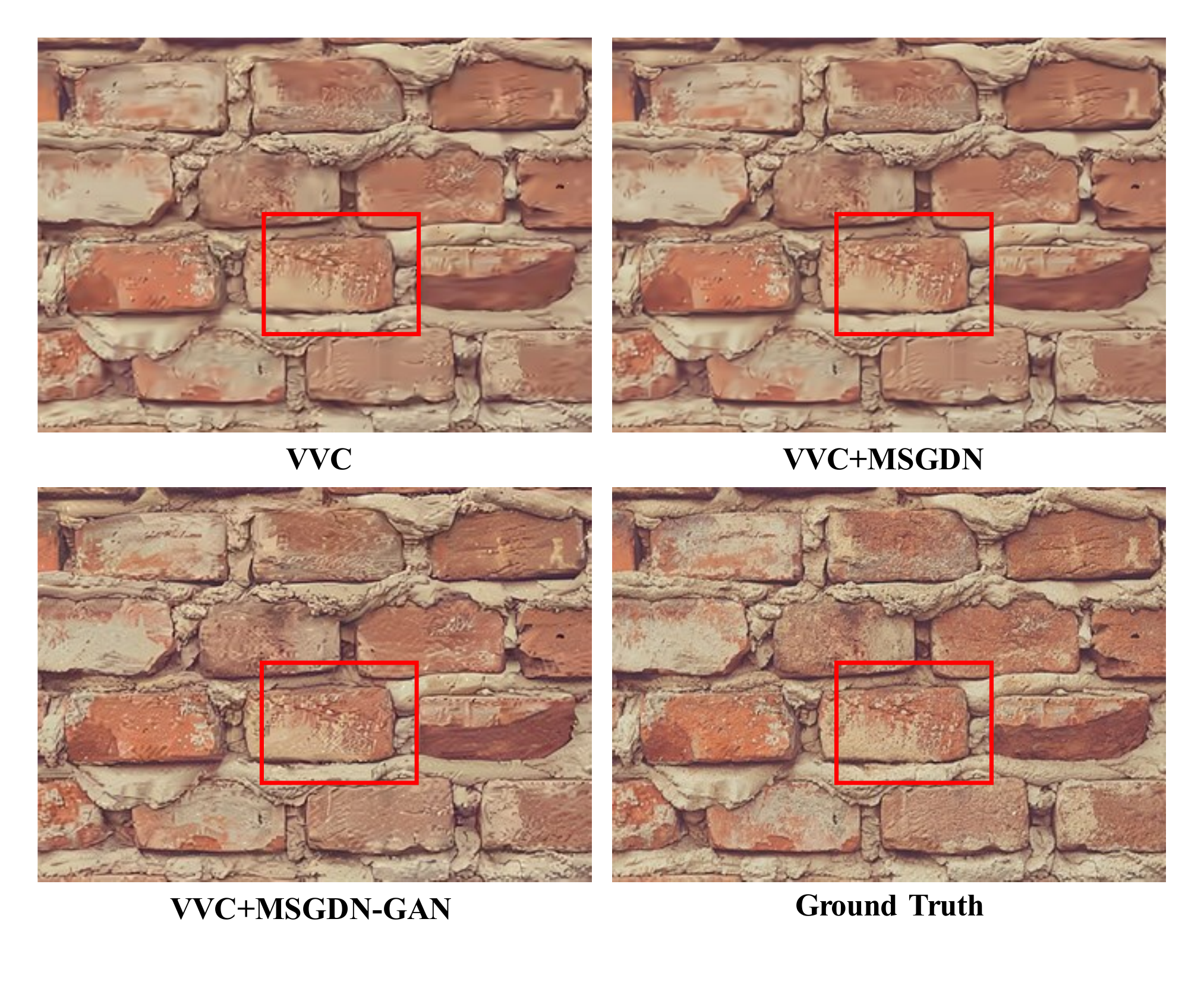}
	\caption{Comparison of image qualities between VVC, VVC+MSGDN, VVC+GAN and Ground Truth, which verify the superiority of our algorithms.}
	\label{fig:compare_3}
\end{figure}

\subsection{Datasets}
For training our MSGDN, we utilize 1633 images from Dataset P (professional) and M (Mobile) and 30000 images which are selected randomly from COCO2014 datasets. Then we transform these images from RGB to YUV444 and use VTM8.0 to compress these images with QP from 37 to 39. Finally, we transform these images from YUV444 to RGB. In this way, we get 31633 pairs of images at each QP.
\subsection{Implementation details}
The MSGDN is implemented based on PyTorch framework with four NVIDIA 1080Ti GPUs. In the process of training,  we set the number of mini-batch as 8 and we utilize Adam optimizer with a initial learning rate of 0.0001. The learning rate will decay by a factor valued 0.5 every 100 epochs. To ensure that the average bpps is 0.15, we conduct image-level bit allocation by selecting images from coded images with QP from 37 to 39. However, in this paper, we only train one MSGDN for these QPs. We find that it cannot bring significant improvement by training different network. 
\subsection{Comparision with traditional methods}
In this section, we compare our approach with traditional coding techniques including JPEG2000, BPG and VTM8.0 respectively. As shown in Fig. \ref{fig:compare_1}, with the same 0.15bpp, the PSNR of our MSGRD is 0.6dB higher than VVC. To compare the subjective quality, we visualize the processed images at about 0.15bpp. As shown in Fig. \ref{fig:compare_3}, the blurring and texture loss are serious in images coded with VVC. However, our MSGDN can remove these artifacts efficiently and our MSGRD-gan achieves the best subjective quality in texture details compared with other methods.

\subsection{Comparison with other post-processing works}
\begin{figure}[htp]
	\centering
	\includegraphics[width=\linewidth]{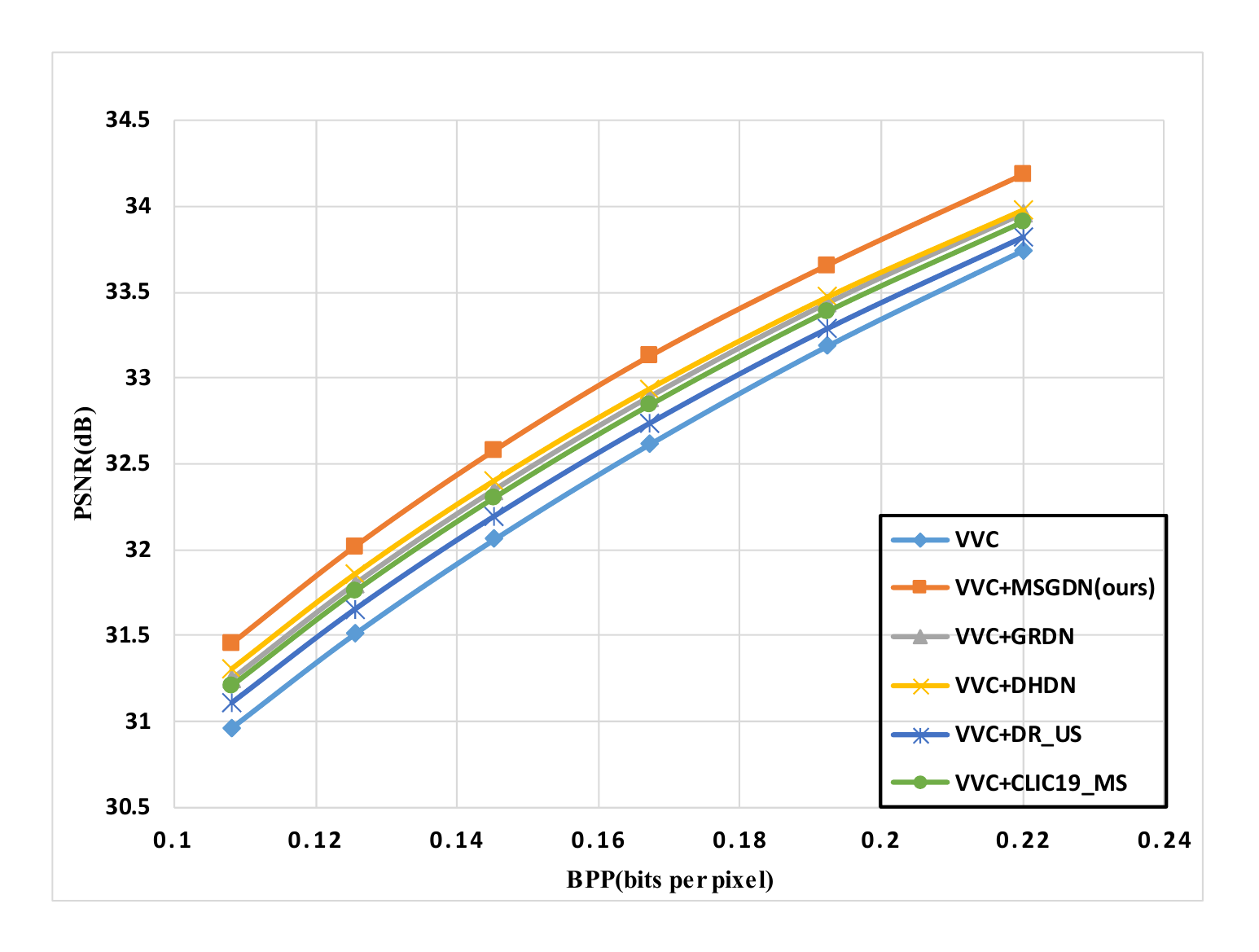}
	\caption{Compression performance on validation sets, compared with other post-process networks.}
	\label{fig:compare_2}
\end{figure}

To further valid the effectiveness of our MSGDN, we re-train a series of post-process network including GRDN \cite{kim2019grdn}, DHDN \cite{park2019densely}, and CLIC2019\_MS \cite{lu2019learned}, DR\_US \cite{liu2019dual}, . As shown in Fig. \ref{fig:compare_2}, our approach achieves the highest PSNR at the point of 0.15bpp, which is 0.3dB than DHDN. 
\vspace{-2mm}
\section{Conclusion}
In this paper, we propose the multi-scale grouped dense  network (MSGDN) as the post-process module of VVC intra coding. By utilizing the multi-scale feature representation and grouped dense block, our MSGDN trained on MSE achieves the PSNR of 32.622 with team "IMC" and "haha" at low bits of 0.15bpp. However, the texture details cannot be reconstructed effectively only with MSE. Therefore, we utilize our MSGDN as a generator, and implement the MSGDN-GAN by combining the loss of RaGAN \cite{jolicoeur2018relativistic} and perceptual loss. Extensive experiments have demonstrated the effectiveness of our MSGDN and MSGDN-GAN. Our MSGDN-GAN achieves better subjective quality in texture details.

\section{Acknowledgement}
This work was supported in part by NSFC under Grant U1908209, 61632001 and the National Key Research and Development Program of China 2018AAA0101400.

{\small
\bibliographystyle{ieee_fullname}
\bibliography{egbib}

\begin{thebibliography}{10}\itemsep=-1pt

\bibitem{balle2016end}
Johannes Ball{\'e}, Valero Laparra, and Eero~P Simoncelli.
\newblock End-to-end optimized image compression.
\newblock {\em arXiv preprint arXiv:1611.01704}, 2016.

\bibitem{chen2019learning}
Zhibo Chen and Tianyu He.
\newblock Learning based facial image compression with semantic fidelity
  metric.
\newblock {\em Neurocomputing}, 338:16--25, 2019.

\bibitem{he2019beyond}
Tianyu He, Simeng Sun, Zongyu Guo, and Zhibo Chen.
\newblock Beyond coding: Detection-driven image compression with semantically
  structured bit-stream.
\newblock In {\em 2019 Picture Coding Symposium (PCS)}, pages 1--5. IEEE, 2019.

\bibitem{jin2018multiscale}
Xin Jin, Runchun Ye, and Zhibo Chen.
\newblock Multiscale progressive image compression network guided by learnable
  just noticeable distortion.
\newblock In {\em 2018 IEEE Visual Communications and Image Processing (VCIP)},
  pages 1--4. IEEE, 2018.

\bibitem{jolicoeur2018relativistic}
Alexia Jolicoeur-Martineau.
\newblock The relativistic discriminator: a key element missing from standard
  gan.
\newblock {\em arXiv preprint arXiv:1807.00734}, 2018.

\bibitem{kim2019grdn}
Dong-Wook Kim, Jae Ryun~Chung, and Seung-Won Jung.
\newblock Grdn: Grouped residual dense network for real image denoising and
  gan-based real-world noise modeling.
\newblock In {\em Proceedings of the IEEE Conference on Computer Vision and
  Pattern Recognition Workshops}, pages 0--0, 2019.

\bibitem{lee2018context}
Jooyoung Lee, Seunghyun Cho, and Seung-Kwon Beack.
\newblock Context-adaptive entropy model for end-to-end optimized image
  compression.
\newblock {\em arXiv preprint arXiv:1809.10452}, 2018.

\bibitem{lee2019hybrid}
Jooyoung Lee, Seunghyun Cho, and Munchurl Kim.
\newblock A hybrid architecture of jointly learning image compression and
  quality enhancement with improved entropy minimization.
\newblock {\em arXiv preprint arXiv:1912.12817}, 2019.

\bibitem{liu2019dual}
Xing Liu, Masanori Suganuma, Zhun Sun, and Takayuki Okatani.
\newblock Dual residual networks leveraging the potential of paired operations
  for image restoration.
\newblock In {\em Proceedings of the IEEE Conference on Computer Vision and
  Pattern Recognition}, pages 7007--7016, 2019.

\bibitem{lu2019learned}
Ming Lu, Tong Chen, Haojie Liu, and Zhan Ma.
\newblock Learned image restoration for vvc intra coding.
\newblock In {\em Proceedings of the IEEE Conference on Computer Vision and
  Pattern Recognition Workshops}, pages 0--0, 2019.

\bibitem{park2019densely}
Bumjun Park, Songhyun Yu, and Jechang Jeong.
\newblock Densely connected hierarchical network for image denoising.
\newblock In {\em Proceedings of the IEEE Conference on Computer Vision and
  Pattern Recognition Workshops}, pages 0--0, 2019.

\bibitem{wang2019integrated}
Mingze Wang, Shuai Wan, Hao Gong, Yuanfang Yu, and Yang Liu.
\newblock An integrated cnn-based post processing filter for intra frame in
  versatile video coding.
\newblock In {\em 2019 Asia-Pacific Signal and Information Processing
  Association Annual Summit and Conference (APSIPA ASC)}, pages 1573--1577.
  IEEE, 2019.

\bibitem{wang2019attention}
Ming-Ze Wang, Shuai Wan, Hao Gong, and Ming-Yang Ma.
\newblock Attention-based dual-scale cnn in-loop filter for versatile video
  coding.
\newblock {\em IEEE Access}, 7:145214--145226, 2019.

\bibitem{wang2018esrgan}
Xintao Wang, Ke Yu, Shixiang Wu, Jinjin Gu, Yihao Liu, Chao Dong, Yu Qiao, and
  Chen Change~Loy.
\newblock Esrgan: Enhanced super-resolution generative adversarial networks.
\newblock In {\em Proceedings of the European Conference on Computer Vision
  (ECCV)}, pages 0--0, 2018.

\bibitem{xue2019attention}
Yuyang Xue and Jiannan Su.
\newblock Attention based image compression post-processing convolutional
  neural network.
\newblock {\em arXiv preprint arXiv:1905.11045}, 2019.

\bibitem{zhang2019learned}
Zhizheng Zhang, Zhibo Chen, Jianxin Lin, and Weiping Li.
\newblock Learned scalable image compression with bidirectional context
  disentanglement network.
\newblock In {\em 2019 IEEE International Conference on Multimedia and Expo
  (ICME)}, pages 1438--1443. IEEE, 2019.

\end{thebibliography}
}

\end{document}